\documentclass[12pt]{iopart}
\pdfoutput=1

\usepackage{graphicx}
\usepackage{iopams}  
\usepackage{url}
\usepackage{multirow}

\begin{document}

\title{A quantitative approach to evolution of music and philosophy}

\author{Vilson Vieira$^1$, Renato Fabbri$^1$, Gonzalo Travieso$^1$, Osvaldo
  N. Oliveira Jr.$^1$, Luciano da
  Fontoura Costa$^1$}

\address{$^1$ Instituto de F\'isica de S\~ao Carlos, Universidade de S\~ao Paulo
  (IFSC/USP)}
\eads{\mailto{vilsonvieira@usp.br}, \mailto{fabbri@usp.br},
  \mailto{gonzalo@ifsc.usp.br}, \mailto{chu@ifsc.usp.br}, \mailto{ldfcosta@gmail.com}}

\begin{abstract}
The development of new statistical and computational methods is
increasingly making it possible to bridge the gap between hard
sciences and humanities. In this study, we propose an approach based
on a quantitative evaluation of attributes of objects in fields of
humanities, from which concepts such as dialectics and opposition are
formally defined mathematically. As case studies, we analyzed the
temporal evolution of classical music and philosophy by obtaining data
for 8 features characterizing the corresponding fields for 7
well-known composers and philosophers, which were treated with
multivariate statistics and pattern recognition methods. A bootstrap
method was applied to avoid statistical bias caused by the small
sample data set, with which hundreds of artificial composers and
philosophers were generated, influenced by the 7 names originally
chosen.  Upon defining indices for opposition, skewness and
counter-dialectics, we confirmed the intuitive analysis of historians
in that classical music evolved according to a master-apprentice
tradition, while in philosophy changes were driven by
opposition. Though these case studies were meant only to show the
possibility of treating phenomena in humanities quantitatively,
including a quantitative measure of concepts such as dialectics and
opposition the results are encouraging for further application of the
approach presented here to many other areas, since it is entirely
generic.
\end{abstract}
\pacs{89.75.Fb,05.65.+b} 

\vspace{2pc}

\noindent{\it Keywords:}\/ Arts, music, musicology, philosophy, pattern recognition, statistics

\maketitle

\section{\label{sec:level1}Introduction}

Philosophy and natural sciences were born together in the 6th century
B.C. within Greek civilization~\cite{Russel}. Logic started with
Aristotle, geometry with Thales and Euclid, arithmetic with
Diophantus, while others created algebra, astronomy and politics, with
all fields being part of a common knowledge
space~\cite{Papineau}. What we know today as modern philosophy and
science were not originally separate, and arts were also present
almost universally in this space, not only as cultural expression. The
harmonic properties of music were of interest to philosophers such as
Pythagoras. Aristotle with his Poetics discussed a dramatic theory for
theatre. In this multidisciplinary space, music was addressed with
mathematics and science with philosophy. Quantitative methods were
applied to explain humanities, while the scientific method had its
origin in philosophy. The segregation of this space into philosophy,
science and arts was inevitable in the scientific revolution, thus
leading to its division into individual, independent areas. Although
specific, these areas are growing in complexity and their domains
overlap. Methods from specific areas are not sufficient to deal with
their complexity.

In this study, we cross the interdisciplinary borders and apply a
quantitative method to philosophy and music, in an attempt to
understand how humanities evolve. While philosophy and music have
their history constantly analyzed and discussed by critics, the nature
of their works --- text documents and music scores --- are difficult
to analyze and interpretation is always subjective. Here we propose a
generic approach to analyze features from any field in a quantitative
manner. The approach, described in the next Section, was then applied
to characterize composers of classical music and philosophers. The
data collected from scores assigned to these composers and
philosophers were treated with statistical and pattern recognition
methods, with the conclusions drawn being then compared to the
literature based on critics of music and philosophy.

\subsection{The Approach}

The approach devised to study the evolution of music and philosophy is
completely generic, and can be applied to any subject. It consists of
the following steps:

\begin{enumerate}
  \item{Given a subject (or area) to be studied, objects belonging to
    this area are identified. Here, philosophers and composers were
    chosen arbitrarily for the fields of philosophy and classical
    music, respectively. In other applications, the choice could be
    objective; for example, in a study of metropolitan areas the
    choice of the objects (cities) would obey a very precise criterion
    (population).}

  \item{A set of attributes is established, which are used to
    characterize the objects. In our case, a few attributes were
    chosen arbitrarily based on well-known characteristics in the
    fields of philosophy and music. In order to make the analysis by
    humans feasible, the number of attributes had to be low. But in
    other types of work, the set could be chosen based on objective
    criteria and the number of attributes could be large.}

  \item{For a quantitative analysis, scores are assigned for each of
    the attributes to each of the objects. In this study, three of the
    authors assigned scores based on their knowledge of the fields
    under study. This can be generalized, with scores assigned with
    objective criteria. Moreover, schemes may be applied to check the
    quality of the assignment, for example using the Kappa
    index~\cite{Cohen1960} to verify agreement among the people who
    assigned the scores.}

  \item{The data generated from the steps above may be sparse and in
    small amount, therefore unsuitable for the application of
    statistical methods. In order to overcome this limitation, we
    introduced a step to verify the robustness of the analysis, which
    consists in generating ``artificial data'' via a bootstrap
    method~\cite{Boos2003}. It is worth noting that in other
    applications, this creation of artificial data may not be
    needed. For instance, we could have taken data for a much larger
    number of philosophers and composers. However, in this paper this
    would hamper the task of manual score assignment, and
    interpretation of the final results would be much more difficult
    (with so many names to be analyzed).}

  \item{Because the study is comparative, metrics must be found to
    identify similarities (or dissimilarities) among the objects, and
    projection methods are applied to visualize the data. There are
    several possibilities for this task, and here we used Pearson
    correlation and principal component analysis~\cite{Costa} to
    analyze the distances among objects. It served to establish a time
    line, with which the evolution of philosophy and music could be
    studied. Furthermore, with this time line it was possible to
    analyze the data in terms of concepts such as dialectics and
    opposition, which were obtained in a quantitative manner.}
\end{enumerate}

With regard to the specific methodology in the present work, we first
identified prominent music composers and philosophers along history,
and established a set of main musical and philosophical
features. Grades were then assigned to each of the composers and
philosophers for all features. The assignment of scores was not
arbitrary, for they were based on research about techniques and styles
used by composers and philosophers. The grades reveal a \emph{tendency
  index} on characteristics of composers or philosophers. For example,
to say that Bach is more contrapuntist than the other composers
corresponds to assigning a higher grade --- e.g.\ 8.0 or 9.0 --- to
the Barroque master and smaller grades to others. We chose a reduced
set of philosophers and composers for the sake of simplicity and
clarity. Then, to avoid statistical bias owing to the small number of
samples, we developed a bootstrap method~\cite{Varian} with which a
larger data set of 1000 new \emph{artificial composers and
  philosophers} were generated, being directly influenced by the
original sample and representing the contemporaries of the
philosophers and composers chosen.

The scores assigned to the characteristics of each composer --- and
philosopher --- define a state vector in its feature space. This
quantification of composers and philosophers allowed the application
of sound quantitative concepts and methods from multivariate
statistics~\cite{Papoulis, Wichern, Therrien} and pattern
recognition~\cite{Duda, Costa}. Correlations between these
characteristic vectors were identified and principal component
analysis (PCA)~\cite{Costa} was applied to represent the music --- and
philosophy --- history as a planar space where development may be
followed as vectorial movements. On this planar space, concepts like
dialectics, innovation and opposition, originally non-quantitative,
can be modeled as mathematical relations between individual states.

It is important to note that application of statistical analysis to
music is not recent. In musicology, statistical methods have been used
to identify many musical
characteristics. Simonton~\cite{Simonton1991829, Simonton1977791} used
time-series analysis to measure the creative productivity of composers
based on their music and popularity. Kozbelt~\cite{Kozbelt01012009,
  Kozbelt01012007} also analyzed productivity, but based on the
measure of performance time of the compositions and investigated the
relation between productivity and versatility. More recent
works~\cite{Kranenburg2004, Kranenburg2007} used machine-learning
algorithms to recognize musical styles of selected compositions.

In contrast to the studies above, we are not interested in applying
statistical analysis to music but on characterizing composers by
identification of scores based on their styles. On the other hand,
automatic information retrieval --- that has been applied to music ---
is not common in philosophy. The method proposed here is a way to
analyze both fields independently of the nature of their works ---
i.e.\ music pieces and textual documents --- but based on a
well-formed opinion of reviewers or critics on these fields and their
analysis of temporal evolution.

\section{Mathematical Description}

Each composer or philosopher and their characteristics
(i.e.\ opposition and skewness) are defined for each pair of
subsequent composers or philosophers along time. Therefore, the choice
of composers and philosophers is crucial for the time-evolution
analysis.  A sequence $S$ of $C$ music composers and $P$ philosophers
was chosen based on their relevance in each period of classical music
and western philosophy history, respectively. The set of $C$
measurements defined a $C$-dimensional space, henceforth referred to
as the \emph{musical space}. Likewise, we defined a $P$-dimensional
\emph{philosophical space} based on $P$ measurements.

\begin{figure}
  \begin{center}
    \includegraphics[width=0.6\textwidth]{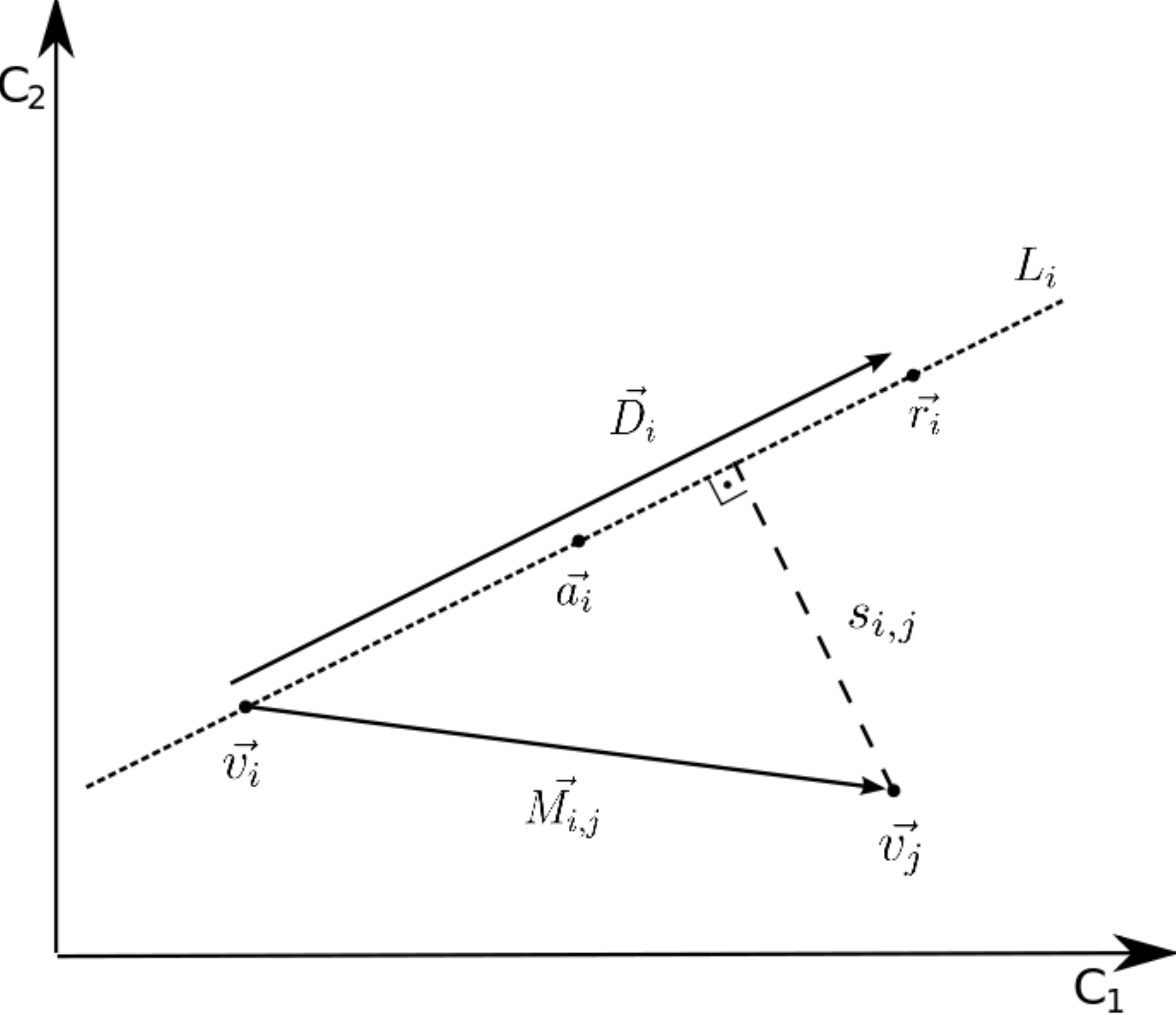}
  \end{center}
  \caption{\it Graphical representation of the measures derived from a 
    \emph{composer state move}.}
  \label{fig.1}
\end{figure}

The vector $\vec{v}_i$ for each composer or philosopher $i$ defines a
corresponding \emph{composer state} in the music space or
\emph{philosopher state} in the philosophy space. For the set of $C$
composers and $P$ philosophers, we defined the same relations
summarized in Table \ref{tab:tableRelations}. Some details about these
relations are worth noting from Figure \ref{fig.1} for composers, and
similar remarks apply to philosophers. Given a set of $C$ composers as
a time-sequence $S$, the \emph{average state} at time $i$ is defined
as $\vec{a}_i$. The \emph{opposite state} is defined as the
``counterpoint'' of a music state $\vec{v}_i$, considering its average
state: everything running along the opposite direction of $\vec{v}_i$
is understood as opposition. In other words, any displacement from
$\vec{v}_i$ along the direction $\vec{r}_i$ is a \emph{contrary move},
and any displacement from $\vec{v}_i$ along the direction $-\vec{r}_i$
is an \emph{emphasis move}. Given a composer state $\vec{v}_i$ and its
opposite state $\vec{r}_i$, the \emph{opposition vector} $\vec{D}_i$
is defined.

\begin{table}
  \caption{\label{tab:tableRelations}Description of mathematical
    relations for each composer or philosopher $i$, $j$ and $k$, given
    a set of $C$ composers or $P$ philosophers as a time-sequence $S$.
    The \emph{opposition index} $W_{i,j}$ quantifies how much a
    composer or philosopher state $i$ is opposed to $j$. The
    \emph{skewness index} $_{i,j}$ quantifies the extent to which the
    new composer or philosopher state $j$ departs from the
    corresponding opposition state. And the \emph{counter-dialectics}
    index quantifies how much a state $k$ could be considered a
    \emph{synthesis} between the \emph{thesis} state $i$ and the
    \emph{antithesis} state $j$.}
  \begin{center}
    \begin{tabular}{ll}
      \\ 
      Average state      & $\vec{a}_i = \frac{1}{i}\sum_{k=1}^i\vec{v}_k.$ 
      \\ \\
      Opposite state     & $\vec{r}_i = \vec{v}_i + 2(\vec{a}_i - \vec{v}_i)$ 
      \\ \\
      Opposition vector  & $\vec{D}_i=\vec{r}_i - \vec{v}_i$ 
      \\ \\
      Composer or \\ 
      philosopher state \\move & $\vec{M}_{i,j} = \vec{v}_j - \vec{v}_i$ 
      \\ \\
      Opposition index  & $W_{i,j} = \frac{\left< \vec{M}_{i,j}, \vec{D}_i\right>}
      {||\vec{D}_i||^2}$ 
      \\ \\
      Skewness index     & $s_{i,j} = \sqrt{\frac{|\vec{v}_i-\vec{v}_j|^2
          |\vec{a}_i-\vec{v}_i|^2 - 
          [(\vec{v}_i-\vec{v}_j) . 
            (\vec{a}_i-\vec{v}_i)]^2}
        {|\vec{a}_i-\vec{v}_i|^2}}$ 
      \\ \\
      Counter-dialectics \\ 
      index              & $d_{i \rightarrow k} = 
      \frac{|\left< \vec{v}_j-\vec{v}_i,\vec{v}_k \right> + 
        \frac{1}{2}\left<\vec{v}_i-\vec{v}_j, \vec{v}_i+\vec{v}_j\right>|}
           {|\vec{v}_j-\vec{v}_i|}$
           \\
    \end{tabular}
    \end{center}
\end{table}

\begin{figure}
  \begin{center}
    \includegraphics[width=0.6\textwidth]{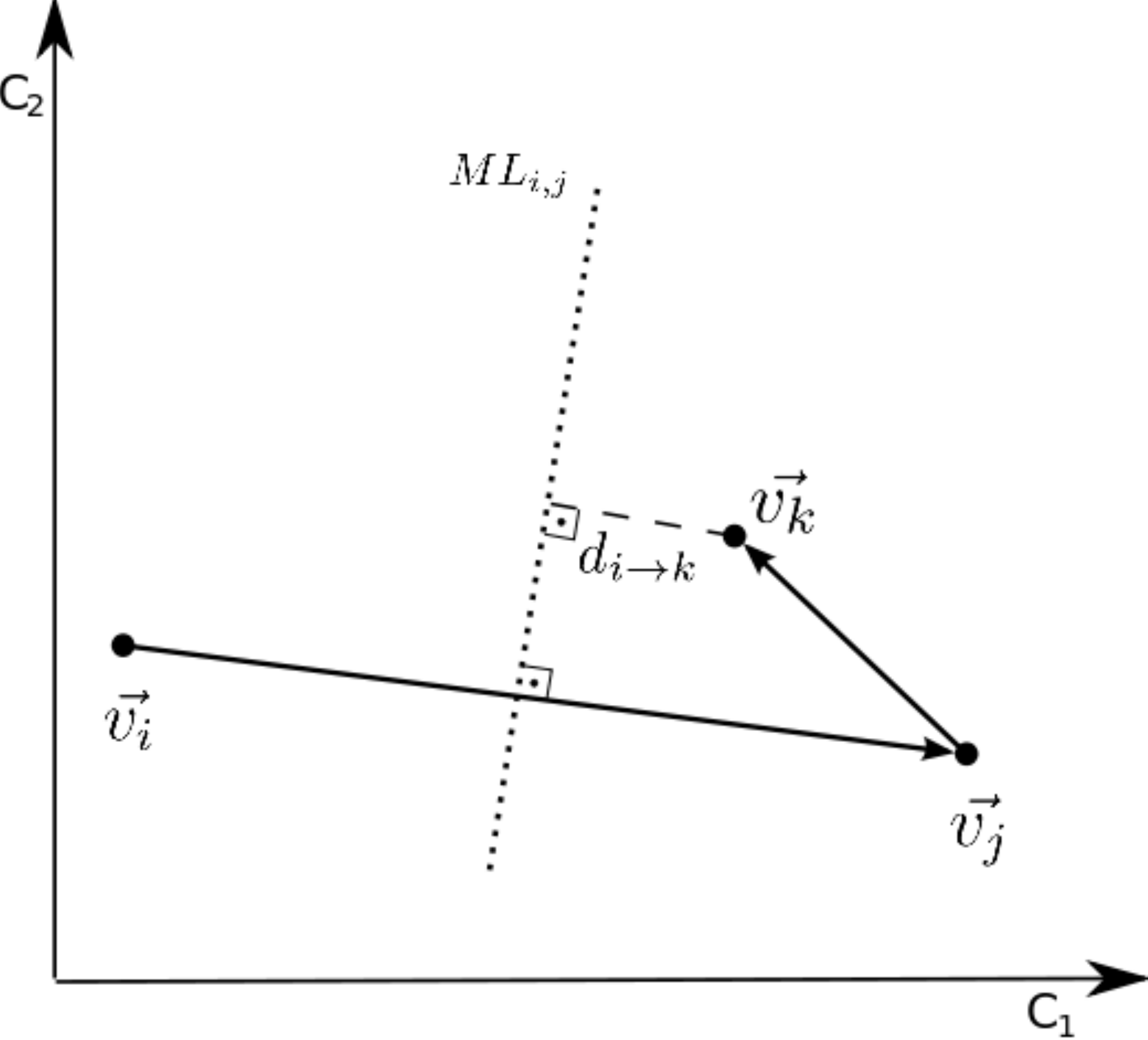}
  \end{center}
  \caption{\it Graphical representation of the quantification of
    dialectics.}
  \label{fig.2}
\end{figure}

For the time-sequence $S$ relations between pairs of composers can be
defined, as follows. The move between two successive \emph{composer
  states} at time $i$ and $j$ corresponds to the $\vec{M}_{i,j}$
vector extending from $\vec{v}_i$ to $\vec{v}_j$. Given the composer
state we can quantify the intensity of opposition by the projection of
$\vec{M}_{i,j}$ along the opposition vector $\vec{D}_i$, normalized,
yielding the \emph{opposition index} $W_{i,j}$. With the same composer
state, the \emph{skewness index} $s_{i,j}$ is the distance between
$\vec{v}_j$ and the line $L_i$ defined by the vector $\vec{D}_i$, thus
quantifying the extent into which the new composer state departs from
the corresponding opposition state. A relationship between a triple of
successive composers can also be defined. Taking $i$, $j$ and $k$ as
the \emph{thesis}, \emph{antithesis} and \emph{synthesis} states, the
\emph{counter-dialectics index} $d_{i \rightarrow k}$ was defined by
the distance between the composer state $\vec{v}_k$ and the middle
line $ML_{i,j}$ defined by the thesis and antithesis, as shown in
Figure \ref{fig.2}. In higher dimensional musical or philosophical
spaces, the middle-hyperplane defined by the points which are at equal
distances to both $\vec{v}_i$ and $\vec{v}_j$ should be used instead
of the middle line $ML_{i,j}$. The proposed equation for
counter-dialectics scales to hyperplanes.  The counter-dialectics
index is used here, instead of a dialectics index, to maintain
compatibility with the use of a distance from point to line as adopted
for the definition of skewness.

\section{Characteristics}

To create the music and philosophy spaces we derived eight variables
corresponding to distinct characteristics commonly found in music
compositions and works by philosophers. While the selected
characteristics cannot be considered a summary of all the relevant
features in music and philosophy history, they are initial indicators
that reveal differences in style of composers and philosophers. We
emphasize that the focus of this work is not on the specific
characteristics used or the scores assigned, which can be disputed,
but on the techniques for a quantitative analysis.

\subsection{Musical Characteristics}

These characteristics are related to basic elements of music ---
melody, harmony, rhythm, timbre, form and
tessitura~\cite{BennettHistory} --- in addition to non-musical issues
like historical events that influenced compositions, such as the
importance of the Church. The eight characteristics are listed
below:\\

{\bf \em{ Sacred - Secular}} ($S$-$S_c$): the sacred or religious
music is composed through religious influence or used for its
purposes. \textit{Masses}, \textit{motets} and hymns, dedicated to the
Christian liturgy, are well-known examples~\cite{Lovelock}. Secular
music has no or minimal relation with religion and includes popular
songs like Italian madrigals and German
\textit{lieds}~\cite{BennettHistory}.

{\bf \em{ Short duration - Long duration}} ($D_s$-$D_l$): compositions
quantified as short duration have a few minutes of execution. Long
duration compositions have at least 20 minutes of execution. The same
criterion was adopted by Kozbelt~\cite{Kozbelt01012009,
  Kozbelt01012007} in his analysis of time execution.

{\bf \em{ Harmony - Counterpoint}} ($H$-$C$): harmony regards the
vertical combination of notes, while counterpoint focuses on
horizontal combinations~\cite{BennettHistory}.

{\bf \em{ Vocal - Instrumental}} ($V$-$I$): compositions using just
vocals (e.g.\ \emph{cantata}) or exclusively instruments
(e.g.\ \emph{sonata}). Note the use of vocals over instruments on
Sacred compositions~\cite{Lovelock}.

{\bf \em{ Non-discursive - Discursive}} ($D_n$-$D$): compositions
based or not on verbal discourse, like programmatic music or Baroque
rhetoric, where the composer wants to ``tell a story'' invoking images
to the listeners mind~\cite{BennettHistory}. Its contrary part is
known as \textit{absolute music} where music was aimed to be
appreciated simply by what it is.

{\bf \em{ Motivic Stability - Motivic Variety}} ($M_s$-$M_v$): motivic
pieces present equilibrium between repetition, reuse and variation of
melodic motives. Bach is noticeable for his \textit{variation} of
motives, contrasting with the constantly inventive use of new
materials by Mozart~\cite{Webern}.

{\bf \em{ Rhythmic Simplicity - Rhythmic Complexity}} ($R_s$-$R_c$):
presence or not of polyrhythms, the use of independent rhythms at the
same time --- also known as \textit{rhythmic
  counterpoint}\cite{BennettHistory} --- a characteristic frequently
found in Romanticism and the works of 20th-century composers like
Stravinsky.

{\bf \em{ Harmonic Stability - Harmonic Variety}} ($H_s$-$H_v$): rate
of tonality change along a piece or its stability. After the highly
polyphonic development in Renaissance, Webern regarded Beethoven as
the composer who returned to the maximum exploration of harmonic
variety~\cite{Webern}.

\subsection{Characteristics in Philosophy}

We derived eight variables corresponding to some of the most recurrent
philosophical issues~\cite{Russel,Papineau,Deleuze2}. Each of these
variables, which defined an axis in the philosophy space, are briefly
described in the following. \\

{\bf \em{ Rationalism - Empiricism}} (\emph{R-E}): the rationalists
claim that the human acquaintance of knowledge/concepts is
significantly independent of sense experience. Empiricists understand
sensory experience as the main way to gain knowledge. Empiricism is in
the origin of the scientific method where knowledge must be based on
sensible observation of the world instead of faith or
intuition. Frequently, rationalists understand the world as affected
by intrinsic properties of the human brain, in contrast to the
empiricist approach where the world would imprint itself onto our
minds (the principle of \emph{``tabula rasa''}). For rationalists,
deduction is the superior method for investigation and privileges the
reason instead of experience as the source of true
knowledge. Historically, Descartes, Espinoza and Leibniz introduced
the rationalism in modern philosophy.

{\bf \em{ Essence - Existence}} (\emph{E-E}): An existence-based
understanding of the world has its basis on the fact that things are
as an existent unit. In Existentialism, the life of an individual is
determined by its self, that constitutes its essence, and not a
predefined essence that defines what is to be human. Essence focuses
on a substance (e.g.\ intellectual) that precedes existence
itself. For essentialists, any specific entity has essential
characteristics necessary for its function and identity. For example,
to have two legs and the ability to run is not an essential
characteristic defining humans, because other animals have the same
characteristics. Plato was one of the first essentialists while
Nietzsche and Kierkegaard were fundamental to Existentialism in the
19th century.

{\bf \em{ Monism - Dualism}} (\emph{M-D}): Dualism requires the
division of the human person into two or more domains, such as matter
and soul. Monism is based on a unique ``category of being''. Plato is
a recognized dualist while his disciple, Aristotle, was a notable
monist.

{\bf \em{ Theocentrism - Anthropocentrism}} (\emph{T-A}): In
theocentrism, God is the most important thing in the universe. The
anthropocentric view has man as prevalent, with Nietzsche being its
main representative. It is important to distinguish theocentrism from
deism, where the figure of God is considered as an abstract entity, as
for Espinoza, but does not play a central role in the universe.

{\bf \em{ Holism - Reductionism}} (\emph{H-R}): Reductionism attempts
to explain the world in terms of simple components and their emerging
properties. Holists focus on the fact that the whole is more than its
constitutive parts.

{\bf \em{ Deductionism - Phenomenology}} (\emph{D-P}): Phenomenology
relies on systematic reflection of consciousness and what happens in
conscious acts. Deductionism is based on deriving conclusions from
axiomatic systems.

{\bf \em{ Determinism - Free Will}} (\emph{D-F}): Free will assumes
that humans make choices, which are not predetermined. Determinism
understands that every event is fatidic, e.g.\, perfectly determined
by prior states.

{\bf \em{ Naturalism - Mechanism}} (\emph{N-M}): Methodological
naturalism is the thinking basis of modern science, i.e. hypotheses
must be argued and tested in terms of natural laws. Mechanism attempts
to build explanation using logic-mathematical processes.

\subsection{Bootstrap method for sampling}
\label{bootstrap}

To eliminate the bias intrinsic in a small sample group, we used a
bootstrap method for generating \emph{artificial composers and
  philosophers} contemporaries of those seven chosen. The bootstrap
routine generated new scores $\vec{r}$, which are not totally random
because they follow a probability distribution that models the
original $n = 7$ scores, given by $p(\vec{r}) = \sum^n_{i=1}
e^{\frac{d_i}{2\sigma^2}}$ where $d_i$ is the distance between a
random score $\vec{r}$ and the original score chart. For each step a
value $p(\vec{r})$ is generated and compared with a random normalized
value, as in the Monte Carlo~\cite{Robert2011} method to choose a set
of samples. These samples simulate new randomized composers and
philosophers score charts --- while preserving the historical
influence of the main 7 original names in each field. Higher values of
$p(\vec{r})$ imply a stronger influence of the original scores over
$\vec{r}$. For the analysis we used 1000 bootstrap samples obtained by
the bootstrap process together with the original scores, taking
$\sigma = 1.1$. Other values for $\sigma$ were used yielding
distributions with bootstrap samples that did not affect the music or
philosophy space substantially.

\section{Results and Discussion}

Memorable composers were chosen as key representatives of classical
music, which we believe had impact on their contemporaries, thus
creating a concise parallel with music history. The chronological
sequence is presented in Table \ref{tab:table0} with each composer
related to his historical period. The same was done for philosophy
where a set of seven philosophers were chosen spanning the period from
Classical Greece until contemporary times, and ordered chronologically
as: Plato, Aristotle, Descartes, Espinoza, Kant, Nietzsche and
Deleuze, as shown in Table \ref{tab:table0b}.

 \begin{table}[ht]
   \caption{\label{tab:table0} Sequence of music composers ordered
     chronologically with the period each represents.}
   \begin{center}
   \begin{tabular}{|l||l|}
     \hline
     Composer        & Movement \\ \hline
     Monteverdi      & Renaissance \\
     Bach            & Baroque \\
     Mozart          & Classical \\
     Beethoven       & Classical $\to$ Romantic \\
     Brahms          & Romantic \\
     Stravinsky      & 20th-century \\
     Stockhausen     & Contemporary\\
     \hline
   \end{tabular}
   \end{center}
 \end{table}

\begin{table}[ht]
  \caption{\label{tab:table0b} Sequence of philosophers ordered
    chronologically with the period each represents.}
  \begin{center}
  \begin{tabular}{|l||l|}
    \hline
    Philosopher     & Era \\ \hline
    Plato           & Ancient \\
    Aristotle       & Ancient  \\
    Descartes       & 17th-century  \\
    Espinoza        & 17th-century  \\
    Kant            & 18th-century  \\
    Nietzsche       & 19th century  \\
    Deleuze         & 20th-century  \\
    \hline
  \end{tabular}
  \end{center}
\end{table}

The quantification of the eight characteristics for music and
philosophy was performed jointly by three of the authors of this
article, based on research of history of music and western
philosophy. The scores shown in Tables \ref{tab:tableAmus} and
\ref{tab:tableAphi} for philosophers and composers, respectively, were
numerical values between 1 and 9. Values closer to 1 reveal the
composer or philosopher tended to the first element of each
characteristic pair, and vice-versa.

\begin{table}[ht]
  \caption{\label{tab:tableAmus} Quantification of eight music
    characteristics for each of the seven composers.}
  \begin{center}
    \begin{tabular}{|l|c|c|c|c|c|c|c|c|}
      \hline
      \footnotesize Composers    & \tiny  $S$-$S_c$ 
      & \tiny  $D_s$-$D_l$ & \tiny  $H$-$C$ & \tiny  $V$-$I$ 
      & \tiny  $D_n$-$D$ & \tiny  $M_s$-$M_v$ & \tiny  $R_s$-$R_c$
      & \tiny  $H_s$-$H_v$  \\
      \hline
   \footnotesize Monteverdi  & 3.0 & 8.0 & 5.0 & 3.0 & 7.0 & 5.0 & 3.0 & 7.0 
   \\
   \footnotesize Bach        & 2.0 & 6.0 & 9.0 & 2.0 & 8.0 & 2.0 & 1.0 & 5.0 
   \\
   \footnotesize Mozart      & 6.0 & 4.0 & 1.0 & 6.0 & 6.0 & 7.0 & 2.0 & 2.0 
   \\
   \footnotesize Beethoven   & 7.0 & 8.0 & 2.5 & 8.0 & 5.0 & 4.0 & 4.0 & 7.0 
   \\
   \footnotesize Brahms      & 6.0 & 6.0 & 4.0 & 7.0 & 4.5 & 6.5 & 5.0 & 7.0 
   \\
   \footnotesize Stravinsky  & 8.0 & 7.0 & 6.0 & 7.0 & 8.0 & 5.0 & 8.0 & 5.0 
   \\
   \footnotesize Stockhausen & 7.0 & 4.0 & 8.0 & 7.0 & 5.0 & 8.0 & 9.0 & 6.0 
   \\
   \hline
    \end{tabular}
  \end{center}
\end{table}

\begin{table}
  \caption{\label{tab:tableAphi} Quantification of eight philosophy
    characteristics for each of the seven philosophers.}
  \begin{center}
    \begin{tabular}{|l||c|c|c|c|c|c|c|c|}
      \hline
      Philosophers & R-E & E-E & M-D & T-A & H-R & D-P & D-F & N-M \\ \hline
      Plato        & 3.0 & 3.5 & 9.0 & 5.0 & 4.5 & 3.5 & 5.0 & 4.5 \\
      Aristotle    & 8.0 & 7.5 & 7.0 & 5.5 & 7.5 & 8.0 & 2.5 & 2.5 \\
      Descartes    & 1.5 & 2.5 & 9.0 & 6.5 & 7.0 & 2.5 & 7.5 & 7.5 \\
      Espinoza     & 8.0 & 2.0 & 1.0 & 5.0 & 2.0 & 3.0 & 1.0 & 1.0 \\
      Kant         & 7.0 & 2.5 & 8.5 & 6.5 & 4.5 & 3.5 & 7.5 & 5.0 \\
      Nietzsche    & 7.5 & 9.0 & 1.0 & 9.0 & 5.0 & 8.0 & 1.0 & 1.5 \\
      Deleuze      & 5.5 & 7.5 & 1.0 & 8.0 & 2.5 & 5.5 & 5.0 & 6.0 \\
      \hline
    \end{tabular}
    \end{center}
\end{table}

This data set defines an 8-dimensional space for music or philosophy
where each dimension corresponds to a characteristic that applies to
all 7 composers or philosophers. Such small data sets are not adequate
for statistical analysis, which could be biased. This is the reason
why we used the bootstrap method for sampling described in
section~\ref{bootstrap}.

For the extended data set after applying the bootstrap method, Pearson
correlation coefficients between the eight characteristics chosen are
given in Table \ref{tab:tableBmus} for composers and in Table
\ref{tab:tableBphi} for philosophers. The coefficient was 0.69 for the
pairs $S$-$S_c$ (Sacred or Secular) and $V$-$I$ (Vocal or
Instrumental), which indicates that sacred music tends to be more
vocal than instrumental. The coefficient 0.56 for the pairs $S$-$S_c$
and $R_s$-$R_c$ (Rhythmic Simplicity or Complexity) also shows that
this genre does not commonly use polyrhythms. A negative coefficient
of -0.33 for the pair $V$-$I$ and $D_n$-$D$ (Non-discursive or
Discursive) indicated that composers who used just voices on their
compositions also preferred programmatic music techniques such as
baroque rhetoric. Strong correlations were also observed for
philosophers. For instance, the Pearson coefficient of $-0.46$ for
\emph{R-E} and \emph{N-M} suggests that rationalists tend to be also
mechanists. An even stronger correlation of $0.74$, now positive, is
observed between \emph{E-E} and \emph{D-P}, i.e. existentialists also
tend to be phenomenologists, as could be expected. Other important
correlations appeared between \emph{D-F} and \emph{N-M} (coefficient =
$0.61$) and between \emph{M-D} and \emph{D-F}, which seems to be
directly implied by religious background.

\begin{table}[ht]
  \caption{\label{tab:tableBmus}Pearson correlation coefficients
    between the eight musical characteristics. The entries with
    absolute values $\geq$ 0.30 have been highlighted.}
  \begin{center}
    \begin{tabular}{|c||c|c|c|c|c|c|c|c|}
      \hline
      -    & \tiny  $S$-$S_c$ & \tiny  $D_s$-$D_l$ & \tiny  $H$-$C$ 
      & \tiny  $V$-$I$ & \tiny  $D_n$-$D$ & \tiny  $M_s$-$M_v$ 
      & \tiny  $R_s$-$R_c$ & \tiny  $H_s$-$H_v$  \\ \hline

      \tiny$S$-$S_c$ 
      & - & -0.2 & -0.06 & \bf{0.69} & -0.18 & 0.19  & \bf{0.56} & -0.16 
      \\
      \tiny$D_s$-$D_l$ 
      & - & - & -0.14 & -0.13 & 0.2 & \bf{-0.48} & -0.2 & \bf{0.37}
      \\
      \tiny$H$-$C$ 
      & - & - & - & -0.23 & 0.26 & 0.05 & \bf{0.46} & 0.03 
      \\
      \tiny$V$-$I$ 
      & - & - & - & - & \bf{-0.33} & 0.17 & \bf{0.42} & -0.06 
      \\
      \tiny$D_n$-$D$ 
      & - & - & - & - & - & -0.3 & 0.02 & -0.22 
      \\
      \tiny$M_s$-$M_v$ 
      & - & - & - & - & - & - & 0.26 & -0.15
      \\
      \tiny$R_s$-$R_c$ 
      & - & - & - & - & - & - & - & -0.02 
      \\
      \tiny$H_s$-$H_v$ 
      & - & - & - & - & - & - & - & - 
      \\
      \hline
    \end{tabular}
    \end{center}
\end{table}

\begin{table}
  \caption{\label{tab:tableBphi}Pearson correlation coefficients
    between the eight philosophical characteristics.  The entries with
    absolute values $\geq$ 0.35 have been highlighted.}
  \begin{center}
    \begin{tabular}{|c||c|c|c|c|c|c|c|c|}
      \hline
      - & R-E & E-E & M-D & T-A & H-R & D-P & D-F & N-M \\ \hline
      R-E & - & \bf{0.37} & -0.23 & 0.15 & 0.1 & \bf{0.46} & -0.27 & \bf{-0.46} 
      \\
      E-E & - & - & \bf{-0.53} & 0.19 & 0.15 & \bf{0.74} & \bf{-0.61} & -0.3 
      \\
      M-D & - & - & - & \bf{-0.43} & \bf{0.41} & -0.3 & \bf{0.35} & 0.01 
      \\
      T-A & - & - & - & - & -0.21 & 0.06  & 0.19 & 0.26 
      \\
      H-R & - & - & - & - & - & 0.32 & -0.22 & -0.25 
      \\ 
      D-P & - & - & - & - & - & - & \bf{-0.63} & \bf{-0.47} 
      \\
      D-F & - & - & - & - & - & - & - & \bf{0.61}
      \\
      N-M & - & - & - & - & - & - & - & - 
      \\
      \hline
    \end{tabular}
    \end{center}
\end{table}

\begin{figure}[htbp]
  \begin{center}
    \includegraphics[width=\textwidth]{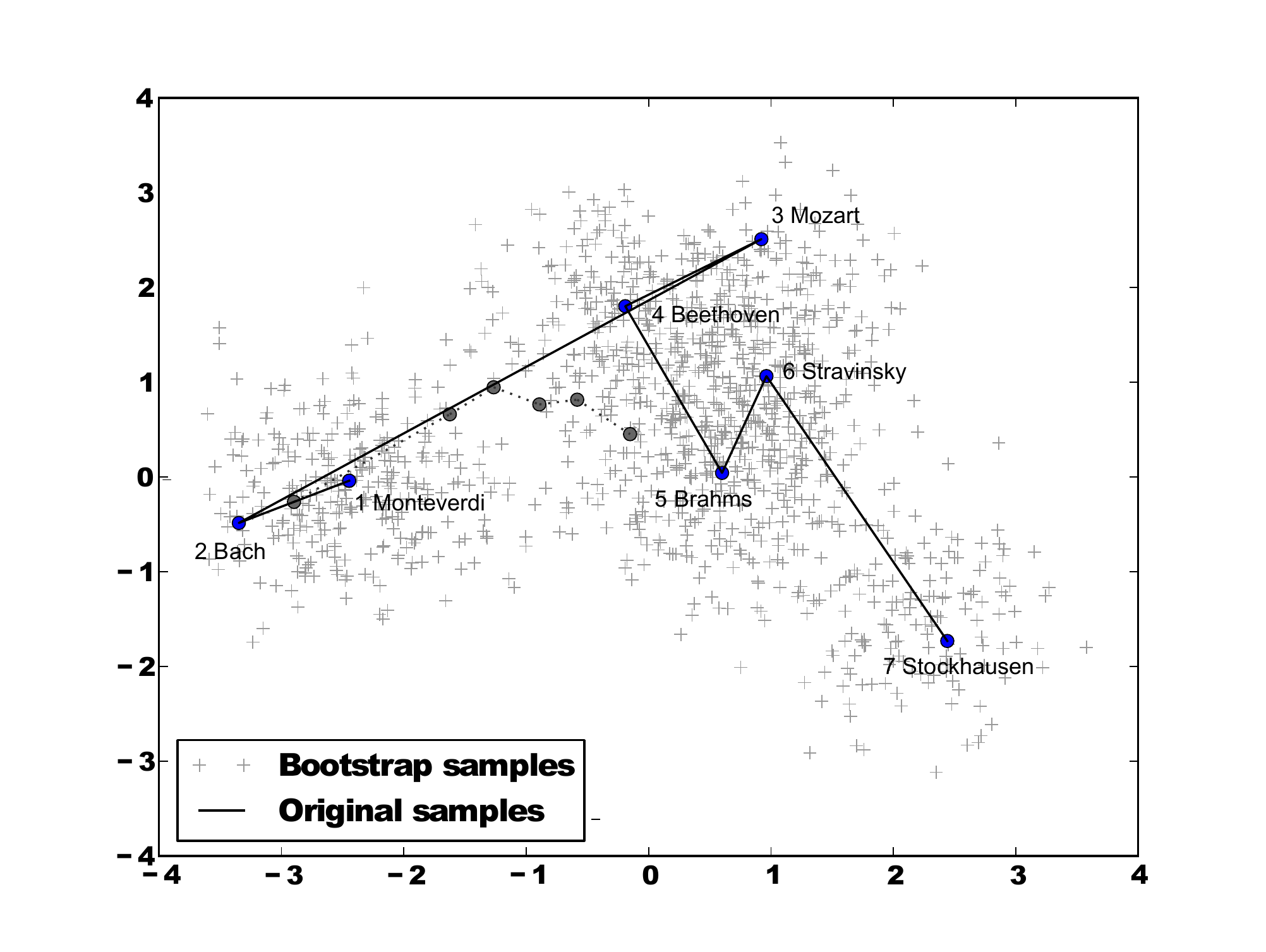}
  \end{center}
  \caption{\it 2-dimensional projected music space.}
  \label{fig:pcamus}
\end{figure}

\begin{figure}
  \begin{center}
    \includegraphics[width=\textwidth]{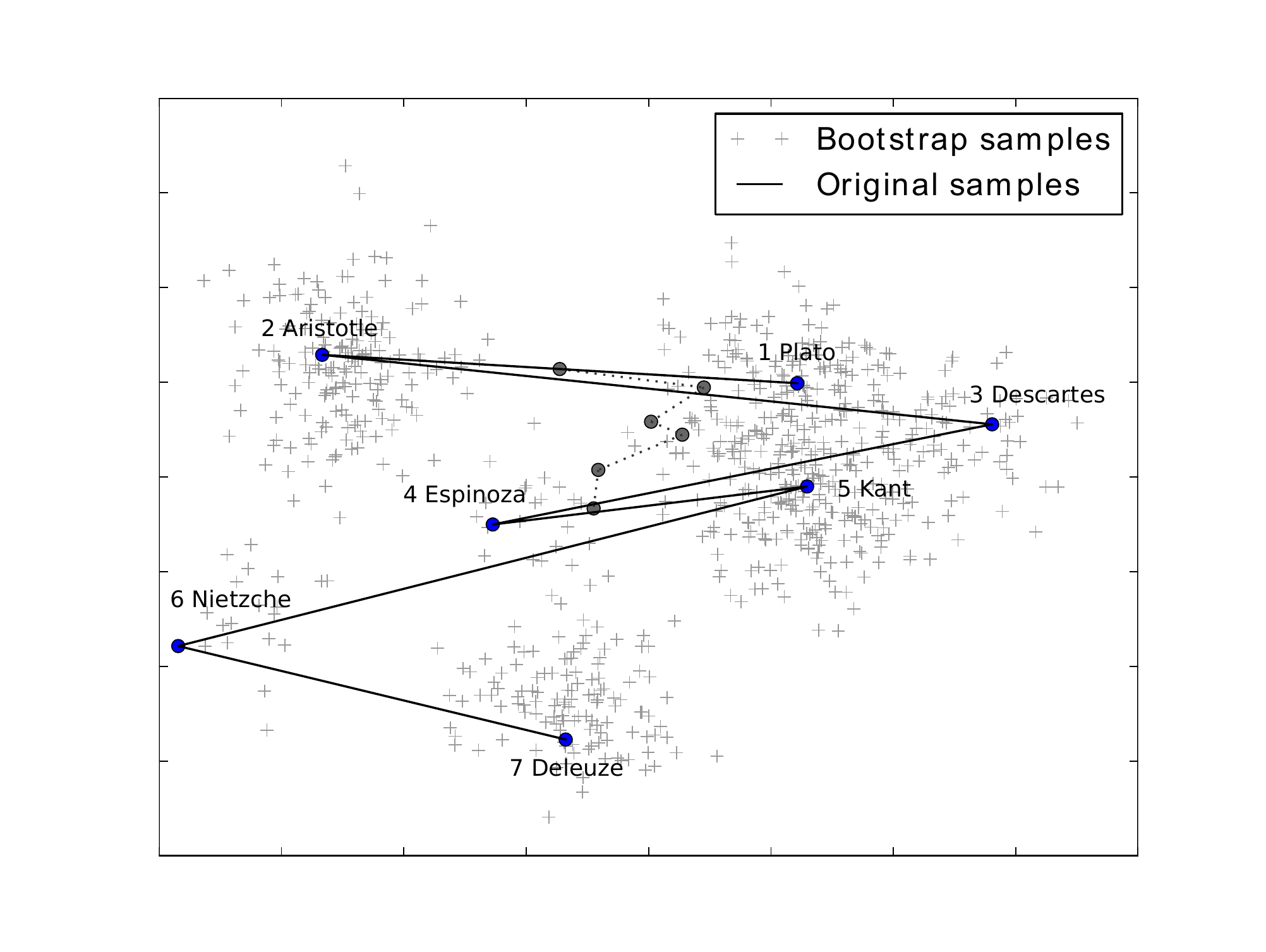}
  \end{center}
  \caption{\it 2-dimensional projected philosophy space.}
  \label{fig:pcaphi}
\end{figure}

The data were analyzed with principal component analysis (PCA), from
which it was found that several of the characteristics contributed to
the variability of the data, as shown in Tables~\ref{tab:Deviatesmus}
and \ref{tab:Deviatesphi} in the Supporting Information. Figures
\ref{fig:pcamus} and \ref{fig:pcaphi} display a 2-dimensional space
with the first two main axes. The arrows follow the time sequence
along with the seven composers and philosophers. Each of these arrows
corresponds to a vectorial move from one composer or philosopher state
to another. For clarity, just the lines of the arrows are
preserved. The bootstrap samples define clusters around the original
composers and philosophers. In subsidiary experiments, we verified
that the results from the bootstrap method were robust. This was
performed by applying 1000 perturbations of the original scores by
adding to each score the values -2, -1, 0, 1 or 2 with uniform
probability. In other words, we tested if scoring errors could be
sufficient to cause relevant effects on the PCA
projections. Interestingly, the values of average and standard
deviation for both original and perturbed positions listed in Tables
\ref{tab:tableDmus} and \ref{tab:tableDphi} in the Supporting
Information show relatively small changes. It is therefore reasonable
to assume that small errors in the scores assigned had no significant
effect on the overall analysis.

Bach is found far from the rest of the composers, which suggests his
key role acknowledged by other great composers like Beethoven and
Webern~\cite{Webern}: ``In fact Bach composed everything, concerned
himself with everything that gives food for thought!''. The greatest
subsequent change takes place from Bach to Mozart, reflecting a
substantial difference in style. There is a strong relationship
between Beethoven and Brahms, supporting the belief by the
\textit{virtuosi} Hans von B\"{u}low~\cite{Bulow} when he stated that
the $1^{st}$ Symphony of Brahms was, in reality, the \textit{$10^{th}$
  Symphony of Beethoven}, appointing Brahms as the true successor of
Beethoven. Stravinsky is near Beethoven and Brahms, presumably due to
his heterogeneity~\cite{BennettHistory, Lovelock}. Beethoven is also
near Mozart, who deeply influenced Beethoven, mainly in his early
works. For Webern, Beethoven was the unique classicist who really came
close to the coherence found in the pieces of the Burgundian School:
``Not even in Haydn and Mozart do we see these two forms as clearly as
in Beethoven. The period and the eight-bar sentence are at their
purest in Beethoven; in his predecessors we find only traces of
them''~\cite{Webern}. It could explain the move of Beethoven in
direction of the Renaissance Monteverdi. Stockhausen is a deviating
point when compared with the others, which could be more even so had
we considered vanguard characteristics --- e.g.\ timbre exploration by
using electronic devices~\cite{Lovelock} --- not shared by his
precursors.

The opposition and skewness indices for each of the six moves among
composers states in Table \ref{tab:tableOImus} indicate that the
movements were driven by small opposition and strong skewness. In
other words, most movements seem to seek innovation rather than
opposition. Furthermore, the counter-dialectics indices in Table
\ref{tab:tableEmus} are smaller than for the philosophers, as will be
discussed later on.

\begin{table}[ht]
  \caption{\label{tab:tableOImus}Opposition and skewness indices for
    each of the six composers states moves}
  \begin{center}
  \begin{tabular}{|c||c|c|}
    \hline
    Musical Move & $W_{i,j}$ & $s_{i,j}$ \\
    \hline
    Monteverdi $\to$ Bach             &   1.0     &  0.      \\
    Bach $\to$ Mozart                 &   1.0196  &  1.9042  \\
    Mozart $\to$ Beethoven            &   0.4991  &  2.8665  \\
    Beethoven $\to$ Brahms            &   0.2669  &  1.7495  \\
    Brahms $\to$ Stravinsky           &   0.4582  &  2.6844  \\
    Stravinsky $\to$ Stockhausen      &   0.2516  &  3.1348  \\
    \hline
  \end{tabular}
  \end{center}
\end{table}

\begin{table}[ht]
  \caption{\label{tab:tableEmus} Counter-dialectics index for each of
    the five subsequent pairs of moves among composers states for the
    8 components.}
  \begin{center}
  \begin{tabular}{|c||c|}
    \hline
    Musical Triple & $d_{i \rightarrow k}$ \\
    \hline
    Monteverdi $\to$ Bach $\to$ Mozart          & 2.0586 \\
    Bach $\to$ Mozart $\to$ Beethoven           & 1.2020 \\
    Mozart $\to$ Beethoven $\to$ Brahms         & 1.0769 \\
    Beethoven $\to$ Brahms $\to$ Stravinsky     & 0.2518 \\
    Brahms $\to$ Stravinsky $\to$ Stockhausen   & 0.2549 \\
    \hline
  \end{tabular}
  \end{center}
\end{table}

As for philosophy, Figure \ref{fig:pcaphi} shows an opposite movement
from Plato to Aristotle, which confirmed the antagonistic view of
Aristotle when compared with Plato, even though Aristotle was his
disciple~\cite{Russel}. Opposition is present along all the moves
among philosophers states. This oscillatory pattern makes it possible
to identify two well defined groups. The first contained Aristotle,
Espinoza, Nietzsche and Deleuze, in the left side of the graph. The
other contained Plato, Descartes and Kant,in the right side.  This
division is consistent with the points of view shared by each group
member. Another two groups are identified in the y-axis, separating
all the philosophers from Nietzsche and Deleuze, who are represented
by the most distant points.

When opposition and skewness indices in Table \ref{tab:tableOIphi} are
considered, all the moves among philosophers states tend to take place
according to a well-defined, intense opposition from the average
state. This was already noticeable in the PCA analysis. An interesting
relationship is the minor opposition and strong skewness between
Nietzsche and Deleuze, suggesting the return to Nietzsche as noted in
the works of Deleuze while considering the vanguard characteristics of
the 20th century philosopher~\cite{Deleuze2}. Espinoza tended toward
Nietzsche, and their similarity was admitted by Nietzsche, by naming
Espinoza his direct precursor~\cite{Russel}. While similar to
Espinoza, Nietzsche presented strong opposition to Kant, consistent
with Nietzsche being the strongest objector to Kant
ideas~\cite{Russel}. Also surprising was the rather small skewness
among most of the moves among philosophers states, which would be
driven almost exclusively by opposition to the current philosopher
state.  The results for dialectics in Table \ref{tab:tableEphi} show a
progressively stronger dialectics among subsequent pairs of moves in
philosophers states.

\begin{table}
  \caption{\label{tab:tableOIphi}Opposition and skewness indices for
    each of the six philosophers states moves.}
  \begin{center}
  \begin{tabular}{|c||c|c|}
    \hline
    Philosophical Move & $W_{i,j}$ & $s_{i,j}$ \\
    \hline
    Plato $\rightarrow$ Aristotle     & 1.0    & 0 \\
    Aristotle $\rightarrow$ Descartes & 0.8740 & 1.1205 \\
    Descartes $\rightarrow$ Espinoza  & 0.9137 & 2.3856 \\
    Espinoza $\rightarrow$ Kant       & 0.6014 & 1.6842 \\
    Kant $\rightarrow$ Nietzsche      & 1.1102 & 2.9716 \\
    Nietzsche $\rightarrow$ Deleuze   & 0.3584 & 2.4890 \\
    \hline
  \end{tabular}
  \end{center}
\end{table}

\begin{table}
  \caption{\label{tab:tableEphi}Counter-dialectics index for each of
    the five subsequent pairs of philosophers states moves.}
  \begin{center}
  \begin{tabular}{|c||c|}
    \hline
    Philosophical Triple & $d_{i \rightarrow k}$ \\
    \hline
    Plato $\rightarrow$ Aristotle $\rightarrow$ Descartes    & 3.0198 \\
    Aristotle $\rightarrow$ Descartes $\rightarrow$ Espinoza & 1.8916 \\
    Descartes $\rightarrow$ Espinoza $\rightarrow$ Kant      & 1.1536 \\
    Espinoza $\rightarrow$ Kant $\rightarrow$ Nietzsche      & 1.1530 \\
    Kant $\rightarrow$ Nietzsche $\rightarrow$ Deleuze       & 0.2705 \\
    \hline
  \end{tabular}
  \end{center}
\end{table}

\begin{figure}[ht]
  \begin{center}
    \includegraphics[width=0.8\textwidth]{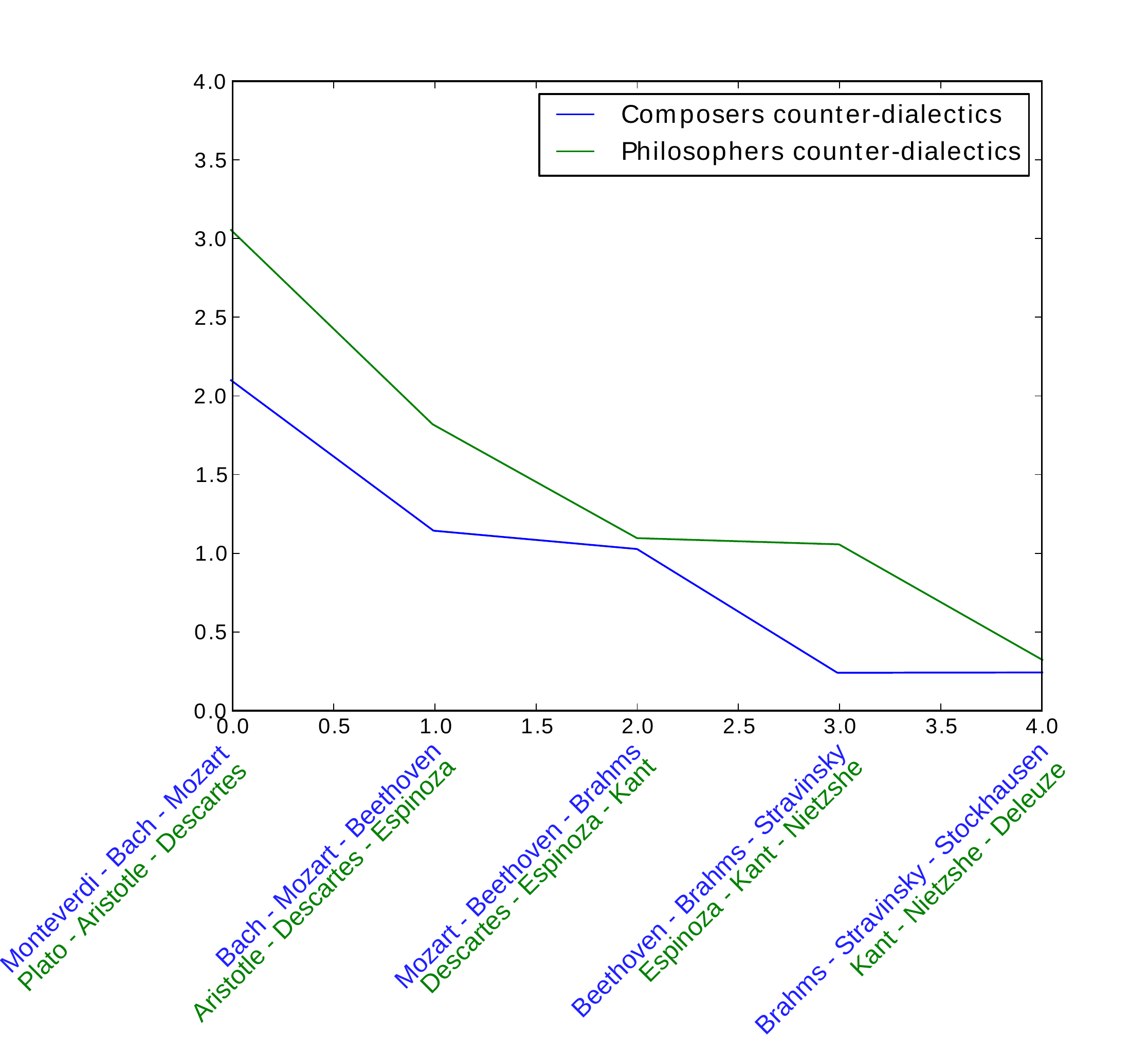}
  \end{center}
  \caption{\it Comparison between composers and philosophers
    counter-dialectics indices}
  \label{fig:comparingdialectics}
\end{figure}

The analysis above indicates distinct characteristics for the
evolution of classical music and philosophy. Philosophers seem to have
developed their ideas driven by opposition ($W_{i,j}$), as shown in
Table \ref{tab:tableOIphi}, while composers tend to be more influenced
by their predecessors according to the dialectics measurements
($1/d_{i \rightarrow k}$). In general, the movements among composers
had minor opposition, thus reflecting the master-apprentice
tradition. There is then a crucial difference in the \textit{memory
  treatment} along the development of philosophy and music: using the
same techniques, we verified that a philosopher was influenced by
opposition of ideas from his direct predecessor, while composers were
influenced by their two predecessors. We can argue that philosophy
exhibits a \textit{memory-1} state, while music presents
\textit{memory-2}, with \textit{memory-N} having $N$ of past
generations that influenced a philosopher or composer. Furthermore,
Figure \ref{fig:comparingdialectics} shows a constant decrease in the
counter-dialectics index, which means a constant return to the origins
for the development of music based on the search for unity. Using the
words of Webern, the search for the ``comprehensibility'' but always
influenced by their old masters.

\begin{figure}[ht]
        \begin{center}
          \includegraphics[width=0.8\textwidth]{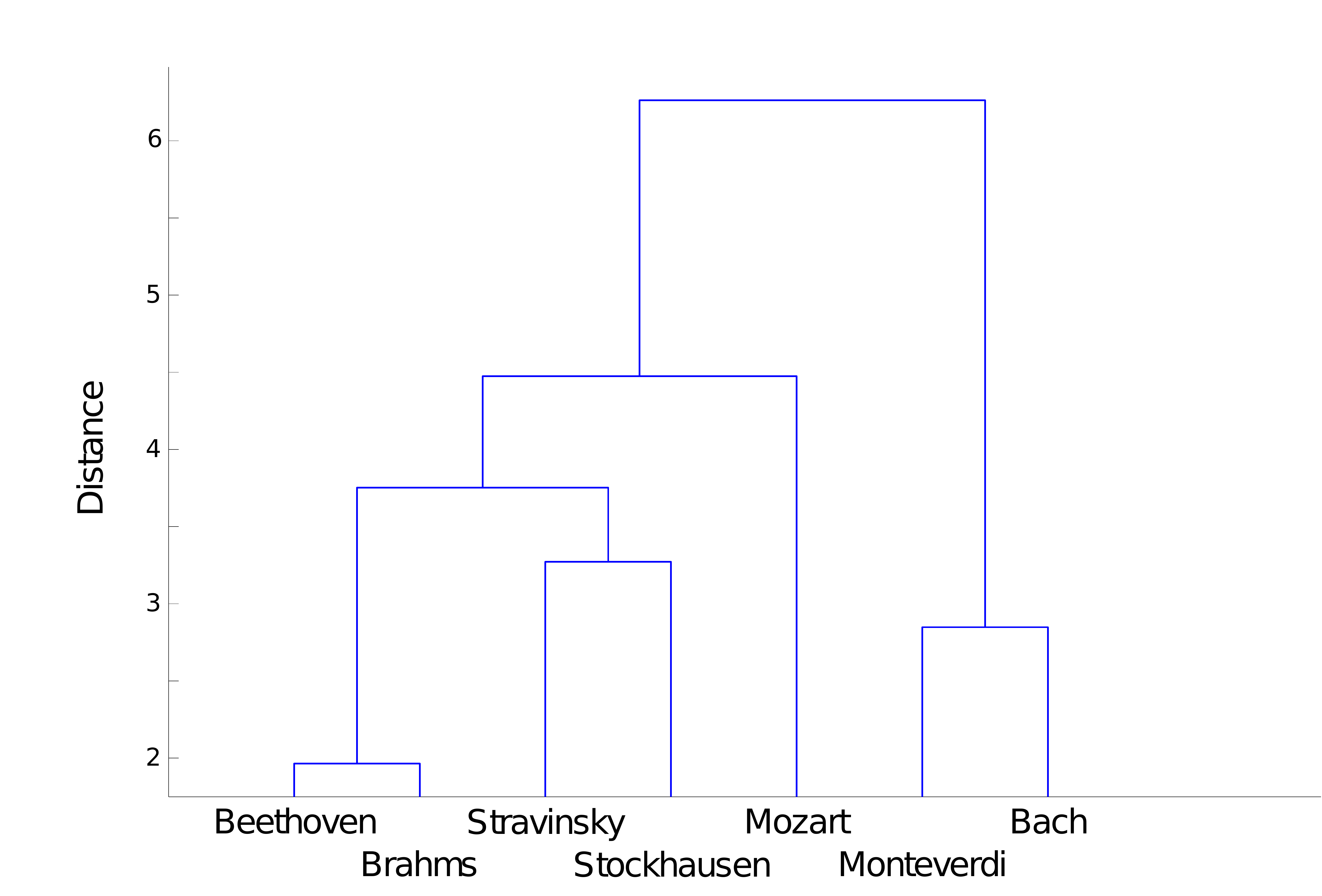}
        \end{center}
        \caption{\it Wards hierarchical clustering of the seven composers.}
        \label{fig:dendrogrammus}
\end{figure}

\begin{figure}
        \begin{center}
                \includegraphics[width=0.8\textwidth]{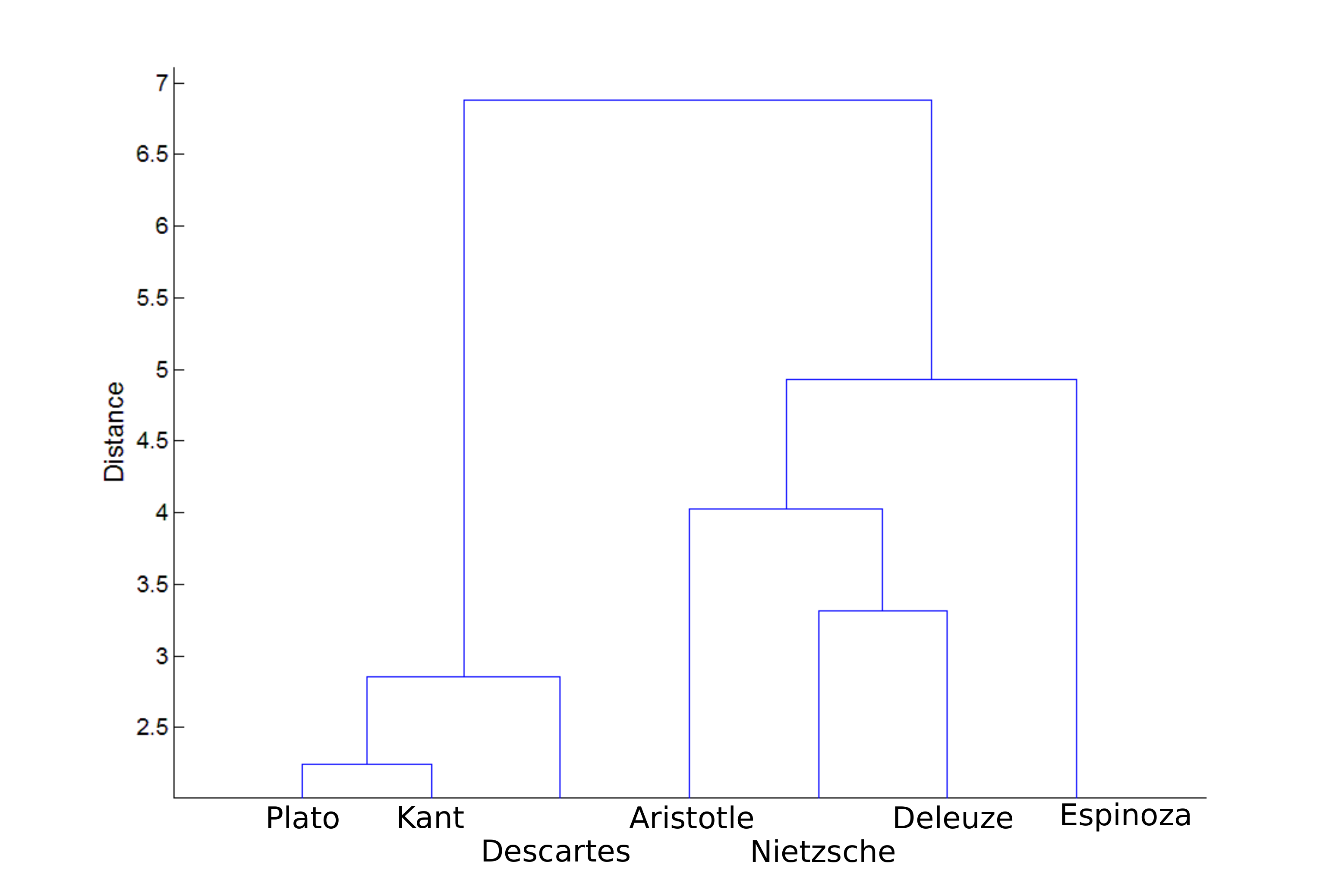}
        \end{center}
        \caption{\it Wards hierarchical clustering of the seven
          philosophers considering all the eight features.}
        \label{fig:dendrogramphi}
\end{figure}

The comparison between composers and philosophers is complemented with
the Wards hierarchical clustering~\cite{Ward}, which clusters the
original scores taking into account their distance. The generated
dendrogram in Figure \ref{fig:dendrogrammus} shows composers according
to their similarity, while the corresponding dendrogram for
philosophers is shown in Figure \ref{fig:dendrogramphi}. These
dendrograms are consistent with the previous observations. For
example, Beethoven and Brahms are close, reflecting their
heritage. Stravinsky and Stockhausen form another cluster, while
Mozart appears on its own, like Bach and Monteverdi. Both relations
were also present in the planar space in Figure \ref{fig:pcamus}.

\section{Concluding Remarks}

An approach has been proposed which allows for a quantitative
assessment of features from any given subject or area. For classical
music and philosophy investigated here, a time line could be
established based on the characteristics of composers and
philosophers. The quantitative analysis involved the collection of
data for these characteristics, extended with a bootstrap method to
yield robustness to the statistical analysis, and the use of
multivariate statistics. Also important was the establishment of
indices for opposition, skewness and counter-dialectics. Though the
emphasis of our work has been on the method of analysis, the
interpretation of the data was already sufficient to draw conclusions
that confirm intuition and subjective analyses of classical music and
philosophy.

For instance, the results pointed to the development in music
following a dichotomy: while composers aim at innovation, creating
their own styles, their technique is based on the work of their
predecessors, in a master-apprentice tradition. Indeed, in the history
of music, composers developed their own styles with a continuous
search for coherence or unity. In the words of Anton
Webern~\cite{Webern}, ``[...] ever since music has been written most
great artists have striven to make this unity ever clearer. Everything
that has happened aims at this [...]''. A frequent inheritance of
style can be identified from one composer to another as a gradual
development from its predecessor, contrasting with the necessity for
innovation. Quoting Lovelock: ``[...] by experiment that progress is
possible; it is the man with the forward-looking type of mind [...]
who forces man out of the rut of `what was good enough for my father
is good enough for me'.''~\cite{Lovelock}. In contrast, our
quantitative analysis confirmed that philosophy appears to exhibit a
well-defined trend in innovation: unlike music, the quest for
difference seems to drive philosophical changes as expressed by Gilles
Deleuze~\cite{Deleuze}. According to Ferdinand de Saussurre's
principle~\cite{Saussure}, concepts (words) tend to be different in
the sense of meaning distinct things.  The paradigm of difference is
particularly important because it is related to the own dynamics of
philosophical evolution.

Again emphasizing that our aim was to propose a generic, quantitative
method, we highlight some limitations of the specific analysis made
here for music and philosophy. For the scores and choice of main
characteristics in music and philosophy were largely arbitrary and
could be disputed. Nevertheless, the perturbation analysis performed
in this work suggests that the effect of non-systematic errors in
assigning the scores does not seem to be critical and has little
overall impact on the conclusions drawn. Most importantly, the formal
quantitative methodology described here may be combined with other
methods, including information retrieval and natural language
processing, to investigate other issues in humanities. For example,
still connected with the present work, it can be adapted to the
investigation of musical and philosophical schools, individual pieces
(e.g.\ music suites or books), or even the contributions from the same
composers or philosophers along distinct periods of time. Obviously,
this methodology can also be applied to other areas such as poetry,
cinema and science.
 
\section{Acknowledgments}

Luciano da F. Costa thanks CNPq (308231/03-1) and FAPESP (05/00587-5)
for sponsorship. Gonzalo Travieso thanks CNPq (308118/2010-3) for
sponsorship. Vilson Vieira is grateful to CAPES.

\appendix

\section{A Brief Explanation of Principal Component Analysis (PCA)}

PCA is a dimensionality reduction procedure performed through rotation
of axes.  It operates by concentrating dispersion/variance along the
first new axes, which are referred to as the principal components. The
technique consists in finding the eigenvectors and eigenvalues of the
covariance matrix of the corresponding random vectors (i.e.\ the
vectors associated with each philosophical state). The eigenvalues
correspond to the variances of the new variables.  When multiplied by
the original feature matrix, the eigenvectors yield the new random
variables which are fully uncorrelated. For a more extensive
explanation of PCA, please refer to~\cite{Costa} and references
therein.

\section{Supporting Information}

Tables \ref{tab:Deviatesmus} and \ref{tab:Deviatesphi} show the
normalized weights of the contributions of each original property on
the eight axes for composers and philosophers. Most of the
characteristics contribute almost equally in defining the axes.

\begin{table}[ht]
  \caption{\label{tab:Deviatesmus}Percentages of
    the contributions from each musical characteristic on the eight
    new main axes.}
  \begin{center}
  \begin{tabular}{|c||c|c|c|c|c|c|c|c|}
    \hline
    Musical         & \multirow{2}{*}{$C_1$} & \multirow{2}{*}{$C_2$} 
    & \multirow{2}{*}{$C_3$} & \multirow{2}{*}{$C_4$} & \multirow{2}{*}{$C_5$}
    & \multirow{2}{*}{$C_6$} & \multirow{2}{*}{$C_7$} & \multirow{2}{*}{$C_8$}
    \\
    Charac. & & & & & & & & \\
    \hline
    $S$-$S_c$   
    &  19.78  &   4.04  & 10.38 & 10.60 &  17.55  &  36.60  &  4.41 &  0.63 \\
    $D_s$-$D_l$            
    &  13.63  &   9.21  & 19.17 &  3.55 &   3.13  &   1.65  & 25.55 & 24.05 \\
    $H$-$C$                
    &   1.44  &  26.62  & 8.26 & 13.97 &  21.71  &   7.76  & 13.98 & 12.20 \\
    $V$-$I$                
    &  18.35  &  12.82  & 9.29 &  8.02 &   9.37  &  40.95  &  2.12 &  2.03 \\
    $D_n$-$D$              
    &   6.31  &  10.73  & 15.48 & 26.29 &  4.04  &   1.86  & 25.29 &  2.35 \\
    $M_s$-$M_v$            
    &  16.94  &  13.28  & 15.03 &  4.84 &  32.25  &  1.70  &  2.62 &  4.37 \\
    $R_s$-$R_c$            
    &  14.13  &   3.26  & 15.58 & 13.80 &   7.48  &  1.88  &  1.36 & 35.99 \\
    $H_s$-$H_v$            
    &   9.38  &  20.00  &  6.75 & 18.88 &   4.45  &  7.56  & 24.62 & 18.36 \\
    \hline
  \end{tabular}
  \end{center}
\end{table}

\begin{table}
  \caption{\label{tab:Deviatesphi}Percentages of
    the contributions from each philosophical characteristic to the four 
    new main axes.}
  \begin{center}
  \begin{tabular}{|c||c|c|c|c|c|c|c|c|}
    \hline
    Philos.      & \multirow{2}{*}{$C_1$} & \multirow{2}{*}{$C_2$} 
    & \multirow{2}{*}{$C_3$} & \multirow{2}{*}{$C_4$} & \multirow{2}{*}{$C_5$} 
    & \multirow{2}{*}{$C_6$} & \multirow{2}{*}{$C_7$} & \multirow{2}{*}{$C_8$}
    \\
    Charac. & & & & & & & & \\
    \hline
    R-E & 13.35   &  1.81 & 13.11 & 31.12 & 15.06 & 10.99 &  8.20 &  3.56 \\
    E-E & 18.13   &  7.88 &  2.10 & 14.68 & 11.41 &  5.31 &  7.98 & 33.50 \\
    M-D & 10.07   & 24.59 & 10.71 &  2.91 &  0.69 & 19.39 & 20.95 &  2.91 \\
    T-A &  1.08   & 24.18 & 21.51 &  0.68 & 23.72 &  6.31 &  7,27 &  2.06 \\
    H-R &  5.90   & 21.49 & 22.29 & 14.90 &  6.35 & 18.81 &  8.69 &  2.05 \\
    D-P & 18.99   &  1.29 &  6.85 &  8.41 &  9.67 & 20.50 & 11.47 & 24.93 \\
    D-F & 14.57   & 13.67 &  8.37 & 18.93 & 21.12 &  9.06 & 12.77 & 14.98 \\
    N-M & 17.86   &  5.06 & 15.03 &  8.33 & 11.96 &  9.59 & 22.62 & 15.97 \\
    \hline
  \end{tabular}
  \end{center}
\end{table}

\begin{table}
  \caption{\label{tab:tableDmus}Averages and standard deviations
    for each composer and for the 8 eigenvalues.}
  \begin{center}
  \begin{tabular}{|c||c|c|}
    \hline
    Composers & $\mu_{\Delta}$ & $\sigma_{\Delta}$ \\
    \hline
    Monteverdi     & 3.7347 & 0.8503 \\
    Bach           & 5.3561 & 0.9379 \\
    Mozart         & 4.4319 & 0.8911 \\
    Beethoven      & 3.4987 & 0.7851 \\
    Brahms         & 3.0449 & 0.6996 \\
    Stravinsky     & 3.6339 & 0.7960 \\
    Stockhausen    & 4.2143 & 0.9029 \\
    \hline \hline
    Eigenvalues & $\mu_{\Delta}$ & $\sigma_{\Delta}$ \\
    \hline
    $\lambda_1$ &  -0.1759 & 0.0045 \\
    $\lambda_2$ &  -0.0638 & 0.0026 \\
    $\lambda_3$ &  -0.0411 & 0.0021 \\
    $\lambda_4$ &  -0.0144 & 0.0019 \\
    $\lambda_5$ &   0.0578 & 0.0021 \\
    $\lambda_6$ &   0.0736 & 0.0023 \\
    $\lambda_7$ &   0.0080 & 0.0027 \\
    $\lambda_8$ &   0.0835 & 0.0030 \\
    \hline
  \end{tabular}
  \end{center}
\end{table}

\begin{table}
  \caption{\label{tab:tableDphi}Average and standard deviation for
    each philosopher and for the 8 eigenvalues.  }
  \begin{center}
  \begin{tabular}{|c||c|c|}
    \hline
    Philosophers & $\mu_{\Delta}$ & $\sigma_{\Delta}$ \\
    \hline
    Plato          & 3.3263 & 0.7673 \\
    Aristotle      & 4.0896 & 0.8930 \\
    Descartes      & 4.3081 & 0.9225 \\
    Espinoza       & 4.9709 & 0.9131 \\
    Kant           & 3.2845 & 0.7749 \\
    Nietzsche      & 5.3195 & 0.9797 \\
    Deleuze        & 4.0990 & 0.8970 \\
    \hline \hline
    Eigenvalues & $\mu_{\Delta}$ & $\sigma_{\Delta}$ \\
    \hline
    $\lambda_1$ & -0.2618 & 0.0068 \\
    $\lambda_2$ & -0.0976 & 0.0035 \\
    $\lambda_3$ &  0.0154 & 0.0025 \\
    $\lambda_4$ &  0.0212 & 0.0024 \\
    $\lambda_5$ &  0.0697 & 0.0026 \\
    $\lambda_6$ &  0.0807 & 0.0030 \\
    $\lambda_7$ &  0.0877 & 0.0032 \\
    $\lambda_8$ &  0.0846 & 0.0036 \\
    \hline
  \end{tabular}
  \end{center}
\end{table}

\bibliographystyle{unsrt}
\section*{References}
\bibliography{phimus}

\end{document}